\documentstyle[11pt]{article}
\begin{document}\openup6pt
\catcode`\@=11

\title{BRANE WORLD INFLATION  WITH SCALAR AND TACHYON  FIELDS}

\author{B.C. Paul\thanks{Email: 
bcpaul@iucaa.ernet.in }  and  Dilip Paul\thanks{Permanent address : Khoribari High School (H.S.), Khoribari, Dist. : Darjeeling }  \\
Physics Department, North Bengal University \\
Dist : Darjeeling, PIN : 734 430, India. 
}

\date{}
\maketitle
\vspace{0.5in}

\begin{abstract}

 We present inflationary models of the early universe in the braneworld scenario considering both scalar field and  tachyon field separately.
The technique of Chervon and  Zhuravlev to obtain inflationary cosmological models
 without restrictions on a scalar field potential is employed here. 
 We note that like scalar field, the inflationary solution obtained here 
 with tachyon field  also does not depend on its potential.   However, 
 unlike scalar field, inflation  with tachyon is obtained for a restricted 
 domains of the field to begin with. We obtain  potentials for which one gets
  inflation  using both scalar field and tachyon field separately. It is found that unlike the scalar field, the tachyonic field inflation sceanario can be realized from $t > t_o$.
 
\end{abstract}


\vspace{2.0cm.}

\pagebreak

{\bf 1. Introduction }\\
It is now generally accepted  that inflation is an essential ingredient in modern cosmology to build cosmological models. Some of the conceptual issues which have no solution in big bang cosmology can be addressed satisfactorily in the framework of  inflationary universe scenario [1].  In the semiclassical theory of gravity the necessary condition for inflation can be  achieved  with matter described by a quantum field. In a scalar field cosmology the potential plays a crucial role in obtaining inflation. 

Recently, Sen [2] shown that the tachyonic condensate in a class of string theories can be described by an effective field, but the Lagrangian for such a tachyonic field is different from that one gets from scalar field.  It is important to 
study cosmology with a tachyonic matter in it as it may play an important role in estimating dark matter in the universe. It is not known definitely what the potential of the tachyonic field is, but there are several attempts to obtain a potential in the context of accelerating universe [3]. Sami [4] obtained inflationary solution with  tachyon in brane model and derived the necessary potential required for the tachyonic slow roll inflation. However, to obtain inflation from a scalar field one has todetermine the slow roll condition for a  successful inflation. Recently,  Chervon [5] and his collaborators  [6] gave an elegant way to rewrite the field equation for a scalar field cosmology which satisfies the condition for inflation automatically in a redefined field. In this paper, we present an inflationary solution of the early universe considering either a scalar field or Sen's tachyonic field [2]. The technique of Zhuravlev and Chervon was employed recently to obtain inflationary solution of the early universe considering tachyon field in GR [7]. We present  here cosmological 
solutions with both scalar field and tachyon field separately  using the above technique and the corresponding 
potentials for the fields are obtained which lead to inflation.\\
 The paper is organized as follows: In sec. 2, the relevant field equations in braneworld model and its general framework to obtain inflationary solutions using Chervon's technique are given. In sec.
 2.1 and 2.2, toy models for the scalar field are discussed. In sec. 3, the field equations with tachyon in braneworld model and its solutions are given. In sec 3.1 and 3.2,  toy models for the tachyon field are discussed. Finally in sec. 4 a brief discussion.
\pagebreak

\vspace*{2pt}
{\bf 2.  Field Equation and Inflationary Solution without Restrictions on scalar field Potential}

\noindent
In the 3+1 dimensional braneworld scenario inspired by the Randall-Sundrum [8] model, the standard Friedmann
equation is modified to [9]:
\begin{equation}
H^2 = \frac{1}{3 M_p^2}\rho\left(1 + \frac{\rho}{2\lambda}\right) + \frac{\Lambda_4}{3} + \frac{\it u_E}{a^4}.
\end{equation} 
where ${\it u_E}$ is an integration constant connected to dark radiation which transmits bulk graviton influence on to the brane, $\lambda$ is the brane tension and $\Lambda_4$ is the effective four-dimensional cosmological constant, $\rho$ represents the matter density. We make the assumption that $\Lambda_4$ is too small to play an important role in the early universe. The "dark radiation" term $\frac{\varepsilon}{a^4}$ is expected to rapidly disappear once inflation has commenced so that we effectively get [9,10]:
\begin{equation}
H^2 = \frac{1}{3 M_p^2}\rho \left(1 + \frac{\rho}{2 \lambda}\right).
\end{equation}
We consider a single minimally coupled homogeneous scalar field which gives 
\begin{equation}
\rho = {\frac{1}{2}}\dot\phi^2 + V(\phi).
\end{equation}
In the braneworld cosmology, at high energy limit where $\rho \gg \lambda$, brane effect becomes important and the modified Friedmann equation (2) can be rewritten as:
\begin{equation}
H^2 \simeq \frac{1}{6 \lambda M_p^2} \left({\frac{1}{2}}\dot\phi^2 + V(\phi)\right)^2.
\end{equation}
The equation of motion of a scalar field propagating on to the brane is
\begin{equation}
 \;
\ddot\phi + 3H\dot\phi + {\frac{dV(\phi)}{d\phi}} = 0.
\end{equation}
where $V(\phi)$ is a potential for the scalar field $\phi $, $H = \frac{d}{dt} \left[ \ln a(t) \right]$ is the Hubble parameter and $a(t)$ is the scale factor of the universe. Using the method adopted in Ref. [5, 6] eqs. (4)-(5) can be rewritten as
\begin{equation}
H = \frac{1}{\sqrt{6 \lambda M_p^2}} W(\phi ). 
\end{equation}
\begin{equation}
3 H  \dot{\phi} = - \frac{d}{d\phi} \left( W(\phi )\right),
\end{equation}
where the total energy density $W(\phi )$ is given by
\begin{equation}
W(\phi) =  V(\phi) + \frac{1}{2}U^2(\phi)
\end{equation}
with $U(\phi) = \dot{\phi} $. Thus the transformed eqs. (6) and (7) looks like that one obtains in the case of a scalar field cosmology where the slow roll conditions are imposed on the field[11 - 13]. Using eqs. (6)-(7) one writes the scale factor in terms of the field and its effective potential(W), which is given by
\begin{equation}
a(t) = a_{o} \;  \exp \left[ - \frac{1}{2\lambda M_p^2} \int \; \frac{W^2(\phi)}{\frac{d}{d \phi}W(\phi)}
d\phi \right].
\end{equation}
The number of e-folds is given by
\begin{equation}
N ( t)  = \int_{t_{o}}^{t_{e}} H dt = \ln \frac{a_{e}}{a_{o}} = - \frac{1}{2\lambda M_p^2} \int_{\phi_{o}}^{\phi_{e}} \frac{W^2(\phi)}{ \frac{d}{d\phi}W(\phi)}d\phi
\end{equation}
where $\phi_{o}$ is the initial value of field when the cosmic inflation begins at $t = t_{o}$ and $\phi_{e}$ is the field value attained when the inflation ends. The functions $U(\phi)$ and $W(\phi)$ are not independent, they are related as :
\begin{equation}
3 W(\phi) \;  U(\phi ) = - (6\lambda M_p^2)^{1/2} \frac{d}{d\phi}W(\phi)
\end{equation}
It is now evident that only one of the functions between $W(\phi)$ and $U(\phi)$ is arbitrary.

Eqs. (4) and (5) can be used to obtain $V$ and $\phi$ as a function of $t$, which are given by 
\begin{equation}
V(t) = (6\lambda M_p^2)^{1/2} \left(\frac{ \dot{a} }{a} \right)\left[1 + \frac{1}{6{(\frac{\dot{a}}{a})}} \; \frac{d}{d t} \left(\ln \frac{d}{d t}\ln a \right)\right]
\end{equation}
\begin{equation}
\phi ( t )  =  \pm \sqrt{ \frac{(6\lambda M_p^2)^{1/2}}{3}} \int \sqrt{- \frac{d}{d t}   \ln  \left( \frac{d}{d t}
\ln a \right)
} \; dt.
\end{equation}
However from eq. (13) the square of the first derivative of the field can also be expressed as
\begin{equation}
 \dot{\phi }^{2} = -\frac{(6\lambda M_p^2)^{1/2}}{3}\left( \frac{\dot{H}}{H}\right).
\end{equation}
From eq. (6), the effective potential can be expressed in term of $H$ as
\begin{equation}
W(\phi) = (6\lambda M_p^2)^{1/2} \; H
\end{equation}
which in turn gives the corresponding  physical potential $V(\phi)$ given by
\begin{equation}
V( \phi ) = (6\lambda M_p^2)^{1/2} \; H \left(1 + \frac{\dot{H}}{6H^{2}}\right)
\end{equation}
which is same as the poptential  $V ( t )$ given in (12). Thus one can express $\phi$ and $V(\phi)$ accordingly. The interesting aspect of the field eqs. (6)-(7) is that these are transformed in such a way that the conditions of inflation are satisfied automatically since  the kinetic term in (6) and $\ddot{\phi}$  term in eq. (7) do not appear in the new set of equations. In the next section we use these equations to obtain cosmological solution.

\vspace{1.0 cm.}

\[
 Toy \; Models \; for \; the \; scalar \; field
\]

{\bf  2.1. Model I :}
\noindent

In the earlier section the set of equations (11) and (13) contain three unknowns, the system of equations may be  solved if one of them is assumed. Let us consider an ansatz for the scalar field
\begin{equation}
\phi = \alpha + \beta t
\end{equation}
where $\alpha$ and $\beta$ are two arbitrary parameters. The Hubble parameter is obtained integrating eq. (14) which is given by
\begin{equation}
H = C \exp\left[-\frac{3\beta^2}{(6\lambda M_p^2)^{1/2}}t\right]
\end{equation}
where $C$ is a constant. On integrating eq. (18) once again one obtains the scale factor which is given by 
\begin{equation}
a(t) = a_{o} \exp\left[-\frac{(6\lambda M_p^2)^{1/2}\; C}{3\beta^2} \exp \left(-\frac{3\beta^2}{(6\lambda M_p^2)^{1/2}}t\right)\right]
\end{equation}
The number of e-folding is given by 
\begin{equation}
{\it  N}  =  -\frac{(6\lambda M_p^2)^{1/2}\; C}{3\beta^2} \left[\exp \left(-\frac{3\beta^2}{(6\lambda M_p^2)^{1/2}}t\right)\right]
_{t_{initial}}^{t_{final}}.
\end{equation}
In this case the effective field potential becomes
\begin{equation}
W(\phi ) = (6\lambda M_p^2)^{1/2}\; C\; \exp \left(-\frac{3\beta}{(6\lambda M_p^2)^{1/2}}(\phi - \alpha)\right)
\end{equation}
The corresponding physical potential $V(\phi)$ which enters in to the original equation can be determined. The potential is
\begin{equation}
V (\phi) = (6\lambda M_p^2)^{1/2}\; C\; \exp \left(-\frac{3\beta}{(6\lambda M_p^2)^{1/2}}(\phi - \alpha)\right) f(\phi)
\end{equation}
where
\begin{equation}
f(\phi) = 1 - \frac{\beta^2}{2(6\lambda M_p^2)^{1/2}\; C}\exp \left(-\frac{3\beta}{(6\lambda M_p^2)^{1/2}}(\phi - \alpha)\right).
\end{equation}

The number of  e-folding in terms of the field becomes
\begin{equation}
{\it N} = -\frac{(6\lambda M_p^2)^{1/2}\; C}{3\beta^2} \left[\exp \left(-\frac{3\beta}{(6\lambda M_p^2)^{1/2}}(\phi_f - \alpha)\right) - \exp \left(-\frac{3\beta}{(6\lambda M_p^2)^{1/2}}(\phi_i - \alpha)\right)\right].
\end{equation}
The inflation, however,  ends when
\begin{equation}
\phi_e = \frac{(6\lambda M_p^2)^{1/2}\; C}{3\beta} \ln\frac {(6\lambda M_p^2)^{1/2}\; C}{3\beta^2} + \alpha
\end{equation}
The inflation ends when the potential attains a value given by
\begin{equation}
V_{end} = \frac{5}{2} \beta^2, \; \; \dot\phi_e = \beta.
\end{equation}

{\bf 2.2  Model II :}
\noindent

Let us consider an ansatz for the scalar field
\begin{equation}
\phi = \alpha \; \ln t
\end{equation}
where $\alpha$ is a parameter ( One can choose $\phi = \alpha \; \ln (t + 1)$ so  that $\phi = 0$ at $t = 0$, but the final conclusion will not affect ). The Hubble parameter is obtained integrating eq. (14) which is given by
\begin{equation}
H = \exp\left[\frac{3\alpha^2}{(6\lambda M_p^2)^{1/2}}\frac{1}{t} + C\right]
\end{equation}
where $C$ is an arbitrary constant. On integrating eq. (28) once again one obtains the scale factor which is given by 
\begin{equation}
a (t) = a_{o} \exp\left[\int\exp\left(\frac{3\alpha^2}{(6\lambda M_p^2)^{1/2}}\frac{1}{t} + C\right)dt\right].
\end{equation}

\vspace*{2pt}
{\bf 3.  Field equations with Tachyon and its solution}

 We now consider tachyon  field in this section. The energy density is given by
In this case the energy density is 
\begin{equation}
\rho(\psi) = \frac{ V(\psi) }{ \sqrt{1 - \dot\psi^{2}}}
\end{equation}
In the high energy limit where $\rho \gg \lambda$, brane effect becomes important and the corresponding Friedmann equations can be rewritten as:
\begin{equation}
H^2 \simeq \frac{1}{6 \lambda M_p^2} \left(\frac{V(\psi)}{\sqrt{1 - \dot\psi^2}}\right)^2.
\end{equation}
The equation of motion of a tachyon field propagating on the brane is
\begin{equation}
\frac{ \ddot{\psi} }{1 - \dot{\psi}^{2}}  + 3 H  \dot{\psi} +  \frac{V'( \psi )}{ V (\psi) } 
= 0
\end{equation}
where $V(\psi)$ is a potential for the tachyon field $\psi $, $H = \frac{d}{dt} \left(\ln a(t) \right)$ is the Hubble parameter and $a(t)$ is the scale factor of the universe. Using the technique adopted by Zhuravlev and Chervon [5, 6 ], eqs. (31) and (32) can be rewritten as
\begin{equation}
H^{2} = \frac{1}{6 \lambda M_p^2} W^2(\psi ), 
\end{equation}
\begin{equation}
3 H  \dot{\psi} = - \frac{d}{d\psi} \left( \ln W(\psi )\right),
\end{equation}
where the effective energy density $W(\phi )$ is given by
\begin{equation}
W(\psi) = \frac{ V(\psi) }{ \sqrt{1 - U^{2}(\psi)}}
\end{equation}
with $U(\psi) = \dot{\psi} $. Thus the eqs. (33) and (34) are expressed in a suitable form which takes the similar form as one requires in the case of slow roll condition for obtaining inflation.
In this case the scale factor of the universe evolves as 
\begin{equation}
a(t) = a_{o} \;  \exp \left[ - \frac{1}{2\lambda M_p^2} \int \; \frac{W^2(\psi)}{\frac{d}{d \psi}\left(\ln W(\psi)\right)}
d\psi \right].
\end{equation}
The number of e-folds is given by
\begin{equation}
N ( t)  = \int_{t_{o}}^{t_{e}} H dt = \ln \frac{a_{e}}{a_{o}} = - \frac{1}{2\lambda M_p^2} \int_{\psi_{o}}^{\psi_{e}} \frac{W^2(\psi)}{ \frac{d}{d\psi}\left(\ln W(\psi)\right)}d\psi
\end{equation}
where $\psi_{o}$ is the initial value of field when inflation begins at $t = t_{o}$ and $\psi_{e}$ is the field when inflation ends. The functions $W(\psi)$ and $W(\psi)$ are related as 
\begin{equation}
3 W(\psi) \;  U(\psi ) = - (6\lambda M_p^2)^{1/2} \frac{d}{d\psi}\left(\ln W(\psi)\right)
\end{equation}
Here only one of the functions between $W(\psi)$ and $U(\psi)$ is arbitrary.
Now we can express eqs. (31) and (32) into a form which can be determined in term of $t$ also, which are
\begin{equation}
V(t) = (6\lambda M_p^2)^{1/2} \left(\frac{ \dot{a} }{a} \right) \left[ \frac{2}{3} + \frac{a\ddot{a}}{3\dot a^{2}} \right]^{1/2},
\end{equation}
\begin{equation}
\psi ( t )  =  \pm { \frac{1}{\sqrt{3}}} \int \sqrt{- \frac{1}{\left(\frac{\dot{a}}{a}\right)} \frac{d}{d t}   \ln  \left( \frac{d}{d t}
\ln a \right)
} \; dt.
\end{equation}
The kinetic term of the field as evident from eq. (31) is
\begin{equation}
\dot{\psi }^{2} = - \frac{\dot{H}}{3 H^2}.
\end{equation}
The effective potential is now can be expressed in terms of Hubble parameter as 
\begin{equation}
W(\psi) = (6\lambda M_p^2)^{1/2} \; H
\end{equation}
and the corresponding  physical potential $V(\phi) $ can be determined. The potential is
\begin{equation}
V( \psi ) = (6\lambda M_p^2)^{1/2} \; H \; \sqrt{1 + \frac{\dot{H}}{3H^{2}}}
\end{equation}
Thus in the case of tachyonic field the technique adopted for a scalar field model may be used to write the field eqs. in brane in such a way that the conditions of inflation are satisfied automatically as the kinetic term in eq.(33) and $\ddot{\psi}$  term in eq. (34) do not appear in the new set of equations.

\vspace{1.0cm.}

\[
 Toy \; Models  \; for \; the \; tachyon \; field
\]

{\bf 3.1 Model I :}
\noindent

In the earlier section we derive the field equations (38) and (40) which have three unknowns. Thus the system of equations may be  solved if one of them is assumed. Let us consider an ansatz for the tachyon field
\begin{equation}
\psi = \alpha + \beta t
\end{equation}
where $\alpha > 1$ and $\beta$ an arbitrary parameter. The Hubble parameter is obtained integrating eq. (41) which is given by
\begin{equation}
H = \frac{1}{3\beta^2t - C}
\end{equation}
where $C$ is a constant. On integrating eq. (45) once again we  obtain  the scale factor which is given by 
\begin{equation}
a(t) = a_{o}\left(3\beta^2t - C)\right)^{\frac{1}{3\beta^2}} .
\end{equation}
The number of e-folding can be expressed in terms of time, which is given by 
\begin{equation}
{\it  N}  =  \frac{1}{3\beta^2}\left[\ln(3\beta^2t - C)\right]_{t_{initial}}^{t_{final}}.
\end{equation}
In this case the effective potential becomes
\begin{equation}
W(\psi ) = (6\lambda M_p^2)^{1/2}\left(\frac{1}{3\beta(\psi - \alpha) - C}\right)
\end{equation}
The corresponding potential $V(\psi)$ is now obtained from eq. (43) in terms of $\psi$.
\begin{equation}
V (\psi) = (6\lambda M_p^2)^{1/2}\left(\frac{\sqrt{1 - \beta^2}}{3\beta(\psi - \alpha) - C}\right)
\end{equation}
which has $1/\psi$ behavior in brane.
The number of  e-folding in terms of field becomes
\begin{equation}
{\it N} = \frac{1}{3\beta^2}\left[\frac{\ln 3\beta(\psi_f - \alpha) - C}{\ln 3\beta(\psi_i - \alpha) - C}\right].
\end{equation}
We note that in this  case    inflation does not depend on the initial value of the inflaton field but there is no end of inflation if  
\begin{equation}
\beta = \pm \frac{1}{\sqrt 3}
\end{equation}
Inflation in this case will end when the potential attains a value
\begin{equation}
V_{end} = \sqrt{\frac{2}{3}}\lambda.
\end{equation}

{\bf 3.2 Model II:}
\noindent

Let us consider another ansatz for the tachyonic  field
\begin{equation}
\psi = \alpha \; \ln t
\end{equation}
where $\alpha$ is a parameter. The Hubble parameter is obtained integrating eq. (41) which is given by
\begin{equation}
H = \frac{1}{\mu - \frac{3 \alpha^{2}}{t}}
\end{equation}
where $\mu$ is an arbitrary constant. On integrating eq. (54) once again one obtains the scale factor which is given by 
\begin{equation}
a (t) = a_{o} e^{\frac{t}{\mu}} \left(\mu t - 3 \alpha^{2} \right)^{\frac{3 \alpha^{2}}{\mu^{2}}}.
\end{equation}
The number of e-folding in terms of time becomes 
\begin{equation}
{\it  N}  =  \left[ \frac{t}{\mu} + \frac{3 \alpha^2}{\mu^{2}} \; \; \ln \left(\mu t - 3 \alpha^{2} \right) \right]_{t_{initial}}^{t_{final}}.
\end{equation}
In this case the effective potential becomes
\begin{equation}
W(\psi ) = \frac{(6\lambda M_p^2)^{1/2}}{\left( \mu  -  3\alpha^{2} e^{-  \frac{\psi}{\alpha} } \right)}.
\end{equation}
The corresponding physical potential $V(\psi)$ now can be expressed as
\begin{equation}
V ( \psi ) = (6\lambda M_p^2)^{1/2} \; \frac{ \sqrt{ 1 - \alpha^{2} e^{- 2 \frac{\psi}{\alpha}}}}{
 \left( \mu -  3 \alpha^{2}\; e^{-  \frac{\psi}{\alpha}} \right)}.
\end{equation}
The number of  e-folding in terms of field becomes
\begin{equation}
{\it N} = \frac{1}{\mu} \left( e^{\psi_{e}/\alpha} - e^{\psi_{o}/\alpha} \right) + 
\ln \left( \frac{\mu e^{\frac{\psi_{e}}{\alpha}} - 3 \alpha^{2} }{\beta e^{\frac{\psi_{o}}{\alpha}} - 3 \alpha^{2}} \right)^{\frac{3 \alpha^{2}}{\mu^{2}}} 
\end{equation}
The inflation epoch ends when
\begin{equation}
\psi_e = \alpha \ln \sqrt{3}\alpha
\end{equation}
with
\begin{equation}
V_{end} = \frac{2  \sqrt{\lambda M_p^2}}{\mu - \sqrt{3}\alpha}.
\end{equation}

It is evident that the tachyon field $\psi$ increases slowly (logarithmically) for $t \geq M_{P}^{- 1}$. However
 $ \dot{\psi}^{2}$ decreases rapidly leading to $V(\psi) \rightarrow constant$ 
 when $\mu \neq \sqrt{3} \alpha$. It is evident that an inflationary 
 scenario is possible for an initial value of the tachyon field $\psi_{o} \geq \alpha \; \ln \sqrt{3}\alpha  \; $ and the tachyonic inflation ends 
 at $\psi_{e} = \alpha \; \ln \sqrt{3}\alpha $. The  tachyonic potential $V (\psi)$  is  restricted as it  becomes imaginary if $\psi < \alpha \; \ln \alpha$.  Thus the potential is regular for a restricted sector of the tachyon field which is different from that of scalar field model as discussed earlier where  $V (\phi)$  is regular everywhere.

\vspace{1.0cm.}

{\bf 4. Discussions }
\noindent

We present cosmological models in braneworld using homogeneous scalar field and tachyonic field separately, We expressed the field equation obtained in  brane  to a suitable form  which gives rise to slow roll inflation. It is found that compared to scalar field inflation, tachyonic inflation is not smooth. We obtain a class of  tachyon potential in which there exists a domain of tachyon field for which inflation permits. It is also noted that for the type I solution where $\psi= \alpha + \beta\; t $, tachyonic inflation is interesting  which exists for $t > t_o =  \frac{ \sqrt{C }}{3 \beta^2}$ and for $t > t_o = \frac{ 3\alpha ^2}{\mu}$ with $\mu > \sqrt{3} \alpha$.

\vspace{1.0cm.}

{\bf Acknowledgement : }

BCP would like to thank  {\bf University Grants Commission, New Delhi} for awarding a Minor Research Project to carry out the work. DP would like to thank {\bf  IUCAA Reference Centre at North Bengal University} for extending support necessary to complete the work.   BCP thankfully  acknowledge  Prof. S. Randjbar-Daemi for awarding  a short term visit to ICTP, Trieste, Italy where the work is completed.

\pagebreak

\end{document}